\documentclass[a4paper]{article}
\usepackage{epsfig,amsmath,amsfonts,url,subfigure,bbm}
\usepackage[text={160mm,237mm}]{geometry}

\def\v#1{\underline{\mathrm{#1}}}  
\newcommand{\tr}{\mathrm{tr}}
\newcommand{\diag}{\mathrm{diag}}

\def\v#1{\mathbf{#1}}  
\newcommand{\triangleq}{=}

\newcommand{\Nr}{{n_r}}
\newcommand{\Nt}{{n_t}}
\renewcommand{\d}{\mathrm{d}}

\newcommand{\Id}{{\bf I}}
\newcommand{\e}{\mathrm{e}}
\title{Maximum Entropy MIMO Wireless Channel Models}
\author{\begin{tabular}{ccc}
Maxime Guillaud &  M\'erouane Debbah & Aris L. Moustakas \\
ftw.            & Eurecom Institute  & University of Athens \\
Vienna, Austria & Sophia-Antipolis, France & Athens, Greece \\
{\tt guillaud@ftw.at}  & {\tt debbah@eurecom.fr}  & {\tt arislm@phys.uoa.gr}
\end{tabular}
}
\date{} 

\begin{document}

\maketitle

\vspace{.5cm}
\section*{Abstract}
In this contribution, models of wireless channels are derived from the maximum entropy principle, for several cases where only limited information about the propagation environment is available.
First, analytical models are derived for the cases where certain parameters (channel energy, average energy, spatial correlation matrix) are known deterministically. Frequently, these parameters are unknown (typically because the received energy or the spatial correlation varies with the user position), but still known to represent meaningful system characteristics. In these cases, analytical channel models are derived by assigning entropy-maximizing distributions to these parameters, and marginalizing them out.
For the MIMO case with spatial correlation, we show that the distribution of the covariance matrices is conveniently handled through its eigenvalues. The entropy-maximizing distribution of the covariance matrix is shown to be a Wishart distribution. Furthermore, the corresponding probability density function of the channel matrix is shown to be described analytically by a function of the channel Frobenius norm.
This technique can provide channel models incorporating the effect of shadow fading and spatial correlation between antennas without the need to assume explicit values for these parameters.
The results are compared in terms of mutual information to the classical i.i.d. Gaussian model.\\

{\bf Keywords:} Maximum Entropy, Multiple Antennas, Wireless Channel Model, Spatial Correlation.
\section{Introduction}

The problem of modelling the characteristics of a wireless transmission channel is crucial to the appropriate design of suitable channel codes. The recent shift to the multiple antennas, or Multiple-Input Multiple-Output (MIMO), paradigm \cite{jerry} and the corresponding need for MIMO channel models, together with the introduction of codes (such as turbo codes \cite{Berrou_etal_turbocodes}) that can operate very close to the channel capacity, has placed the channel models under scrutiny:
initial capacity analyses of MIMO channels assuming i.i.d. Rayleigh fading \cite{telatar_mimo_capacity} were touting promising spectral efficiencies, whereas the importance of correlation between channel coefficients \cite{kyritsi_etal_mimo_correlation} and of the channel matrix rank 
are now understood to be critical parameters.
In order to facilitate channel code development, analytical channel models are a desirable asset. However, most of the available channel models that capture the complex spatial characteristics of the propagation channel (geometry, reflection coefficients, \ldots) are based on ray tracing methods or variations thereof, which model the channel as a superposition of multipath components \cite{Burr_finite_scatterers} and therefore do not lend themselves easily to analysis. Conversely, some analytical models were proposed to address the problem of accurate space correlation modeling by assuming a Rayleigh fading with appropriately designed correlation properties \cite{Weichselberger_model}. See \cite{survey_MIMO_models_JWCN} for a broad round-up of the literature about wireless channel models.\\

In \cite{Debbah_maxent}, Debbah and M\"uller address the question of channel modeling on the basis of statistical inference.
Instead of relying on ad-hoc construction -- based on intuition -- and verification of the models, they propose a constructive method based on the constraints that the model needs to meet. 
The joint probability density function (PDF) of the channel is derived from these constraints, using the maximum entropy (MaxEnt) principle, initially introduced by Jaynes \cite{Jaynes_1957}.
This principle build on the fact that the only consistent way of accounting for ignorance when modelling a random process is to maximize the entropy of the considered process, subject to all known constraints. In this context, consistent modelling is defined as the requirement that independent modellers being given the same set of constraints must obtain identical models. This approach is justified on the basis of avoiding the arbitrary introduction of information (in the form of model characteristics that represent a reduction of its entropy) and that can not be justified by any known constraint.
In the case of channel modelling, the constraints represent available knowledge about the environment or the channel representation itself (e.g. through bounds on amplitude, power...).
See \cite{Jaynes_probability_theory} for a recent overview of the application of maximum entropy methods to inference.\\

In \cite{Debbah_maxent}, the MaxEnt principle is used to derive a joint distribution of the entries of the MIMO channel matrix. The popular Gaussian i.i.d. model is shown to be the entropy-maximizing solution under the sole assumption that the average Frobenius norm of the channel matrix is known (known channel power constraint). However, this model is admittedly simplistic, in particular because of the following two reasons:
\begin{itemize}
\item Measurements have shown that the independence between components, as obtained in \cite{Debbah_maxent} and proposed in numerous models, rarely holds in reality, and that some degree of correlation between the components must be taken into account,
\item Gaussian models constitute good short-term models but their long-term properties are not realistic. More precisely, Gaussian models are known to adequately model the effects of rich scattering, but to neglect the long-term fading effect captured by the fact that the signal strength (represented by the short-term average of the channel Frobenius norm - in the following this quantity is denoted by ``channel energy'') fluctuates.
\end{itemize}
The aim of the present article is to extend the general scope of maximum entropy channel modeling, by amending existing models to address the aforementioned issues. Both points are addresses using the same method: first, a maximum entropy model is derived for the channel, conditioned on the parameter of interest (signal strength or spatial correlation). Then, a maximum entropy distribution for is derived for the parameter of interest itself, and is later marginalized out to obtain the full channel model.\\

This article is structured as follows: first, some notations are introduced in Section \ref{section_notations}. In Section \ref{section_energyconstraints}, a maximum entropy model for the channel energy is proposed, based on the knowledge of the average, and optionally on an upper bound, of the channel energy. The corresponding channel model is obtained by first deriving the distribution of the instantaneous channel realization for a known channel energy, and in a second step by marginalizing out the variable representing the energy using the distribution established previously.
Section \ref{section_spatialcorrelation}, focuses on the spatial correlation properties of frequency-flat fading channels. Specifically, we address the case where the channel is known to have spatial correlation, but the exact characteristics of this correlation are not known. In general, in the absence of knowledge about correlation, application of the MaxEnt principle yields a process with independent components (see \cite{Debbah_maxent}).  Therefore, we first focus on the spatial covariance matrix, and derive the MaxEnt distribution of a general covariance matrix, in both the full-rank and rank-deficient cases. In a second step, we construct the analytical model for the MIMO channel itself, by first deriving the MaxEnt distribution of the channel for a known covariance, and later marginalizing over the covariance matrix, using the distribution of the covariance established previously.
The obtained distribution is shown to be isotropic, and is described analytically as a function of the Frobenius norm of the channel matrix. Finally, Section \ref{section_conclusion} draws some conclusions.\\

\section{Notations and channel model}
\label{section_notations}

Let us consider the multiple-antenna wireless channel with $\Nt$ transmit and $\Nr$ receive antennas. Since we are only concerned with non-frequency selective channels, let the complex scalar coefficient $h_{i,j}$ denote the channel attenuation between transmit antenna $j$ and receive antenna $i$, $j=1\ldots \Nt$, $i=1\ldots \Nr$. Let $\mathbf{H}(t)$ denote the $\Nr\times \Nt$ channel matrix at time $t$.
We recall the general model for a time-varying flat-fading channel with additive noise
\begin{equation}
  \v{y}(t) = \mathbf{H}(t)\v{x}(t) + \v{n}(t), \label{awgn_channelmodel}
\end{equation}
where $\v{n}(t)$ is usually modeled as a complex circularly-symmetric Gaussian random variable (r.v.) with independent identically distributed (i.i.d.) coefficients.
In this article, we focus on the derivation of the fading characteristics of $\mathbf{H}(t)$. When we are not concerned with the time-related properties of $\mathbf{H}(t)$, we will drop the time index $t$, and refer to the channel realization $\mathbf{H}$ or equivalently to its vectorized notation $\v{h} \triangleq \mathrm{vec}(\mathbf{H}) = [h_{1,1}\ldots h_{\Nr,1}, h_{1,2}\ldots  h_{\Nr,\Nt}]^T$.
Let us also denote $N\triangleq \Nr\Nt$ and map the antenna indices into $[1\ldots N]$, \emph{i.e.} denoting equivalently $\v{h}=[h_1\ldots h_N]^T$.\\

\section{Channel energy constraints}
\label{section_energyconstraints}

\subsection{Average channel energy constraint}
\label{previous_results}
In this section, we briefly recall the results of \cite{Debbah_maxent}, where an entropy-maximizing probability distribution is derived for the case where the average energy of a MIMO channel is known deterministically. 
It is obtained by maximizing the entropy $\int_{\mathbb{C}^N} -\log(P(\mathbf{H})) P(\mathbf{H}) \d\mathbf{H}$, where $\d\mathbf{H}\triangleq \prod_{i=1}^N \mathrm{dRe}(h_i) \mathrm{dIm}(h_i)$ is the Lebesgue measure on $\mathbb{C}^N$ ($\mathrm{Re}(\cdot)$ and $\mathrm{Im}(\cdot)$ denoting respectively the real and imaginary parts of a complex number), under the only assumption that the channel has a finite average energy $N E_0$, and the normalization constraint associated to the definition of a probability density, \emph{i.e.}
\begin{equation}
  \int_{\mathbb{C}^N} ||\mathbf{H}||_{F}^2 P(\mathbf{H}) \d\mathbf{H} = N E_0, \quad \textrm{and}    \int_{\mathbb{C}^N} P(\mathbf{H}) \d\mathbf{H} = 1. \label{normintegrals}
\end{equation}
This is achieved through the method of Lagrange multipliers, by writing
\begin{equation}
  L(P) = \int_{\mathbb{C}^N} -\log(P(\mathbf{H})) P(\mathbf{H}) \d\mathbf{H} + \beta \left[1- \int_{\mathbb{C}^N} P(\mathbf{H}) \d\mathbf{H}\right] + \gamma \left[N E_0 - \int_{\mathbb{C}^N} ||\mathbf{H}||_{F}^2 P(\mathbf{H}) \d\mathbf{H}\right]
\end{equation}
where we introduce the scalar Lagrange coefficients $\beta$ and $\gamma$, 
and taking the functional derivative \cite{Fomin_calculus_variations} w.r.t. $P$ equal to zero:
\begin{equation}
  \frac{\delta L(P)}{\delta P} = -\log(P(\mathbf{H}))-1 -\beta -\gamma||\mathbf{H}||_{F}^2 =0.
  \label{derivativeP}
\end{equation}
Eq.~(\ref{derivativeP}) yields $P(\mathbf{H}) = \exp\left( -(\beta+1) -\gamma ||\mathbf{H}||_{F}^2\right)$, and the normalization of this distribution according to (\ref{normintegrals}) finally yields the coefficients $\beta$ and $\gamma$, and the final distribution is obtained as
\begin{equation}
  P_{\mathbf{H}|E_0}(\mathbf{H}) = \frac{1}{(\pi E_0)^{N}} \exp\left( -\sum_{i=1}^{N} \frac{|h_{i}|^2}{E_0}  \right) \label{maxent_Giid}
\end{equation}
Interestingly, the distribution defined by eq.~(\ref{maxent_Giid}) corresponds to a complex Gaussian random variable with independently fading coefficients, although neither Gaussianity nor independence were among the initial constraints. These properties are the consequence, \emph{via} the maximum entropy principle, of the ignorance by the modeler of any constraint other than the total average energy $N E_0$.\\

\subsection{Probabilistic average channel energy constraint}
Let us now introduce a new model for situations where the channel model defined in the previous section applies locally (in time), but where $E_0$ can not be expected to be constant, e.g. due to shadow fading. Therefore, let us replace $E_0$ in eq.~(\ref{maxent_Giid}) by the random quantity $E$, known only through its probability density function (PDF) $P_E(E)$. In this case, the PDF of the channel $\mathbf{H}$ can be obtained by marginalizing over $E$:
\begin{equation}
  P(\mathbf{H}) = \int_{\mathbb{R}^+} P_{\mathbf{H},E}(\mathbf{H},E) \d E  = \int_{\mathbb{R}^+} P_{\mathbf{H}|E}(\mathbf{H}) P_E(E) \d E. \label{unknown_E}
\end{equation}

In order to establish the probability distribution $P_E$, let us find the maximum entropy distribution under the constraints:
\begin{itemize}
  \item $0 \leq E \leq E_{max}$, where $E_{max}$ represents an absolute constraint on the transmit power, or on the amplitude range of the receiver,
  \item its average $E_0 \triangleq \int_{0}^{E_{max}} E P_E(E) \d E$ is known.
\end{itemize}
Applying the Lagrange multipliers method again, we introduce the scalar unknowns $\beta$ and $\gamma$, and maximize the functional
\begin{equation}
  L(P_E) = - \int_0^{E_{max}} log(P_E(E)) P_E(E)\d E + \beta\left[ \int_0^{E_{max}} E P_E(E) \d E - E_0 \right] + \gamma\left[ \int_0^{E_{max}}  P_E(E) \d E -1 \right].
\end{equation}
Taking the derivative equal to zero ($\frac{\delta L(P_E)}{\delta P_E} = 0$) yields $P_E(E) = \exp\left( \beta E - 1 + \gamma \right)$, and the Lagrange multipliers are finally eliminated by solving the normalization equations
\begin{equation}
  \int_0^{E_{max}} E \exp\left( \beta E - 1 + \gamma \right)\d E = E_0, \quad \mathrm{and}\quad \int_0^{E_{max}}  \exp\left( \beta E - 1 + \gamma \right) \d E =1.
\end{equation}
$\beta<0$ is the solution to the transcendental equation
\begin{equation}
  E_{max} \exp(\beta E_{max}) - \left(\frac{1}{\beta} +E_0\right)\left(\exp(\beta E_{max}) -1\right) =0, \label{solution_beta}
\end{equation}
and finally $P_E$ is obtained as the truncated exponential law
\begin{equation}
  P_E(E)=\frac{\beta}{\exp(\beta E_{max})-1} \exp(\beta E), \quad 0 \leq E \leq E_{max}, \quad  0\quad \mathrm{elsewhere}.
\end{equation}
Note that taking $E_{max}=+\infty$ in eq.~(\ref{solution_beta}) yields $\beta=-\frac{1}{E_0}$ and the exponential law $ P_E(E)=E_0 \exp\left(-\frac{E}{E_0} \right)$.

\subsubsection{Application to the SISO channel}
\begin{figure}[htb]
\centering
\subfigure[PDF of amplitude $r$]{\epsfig{figure=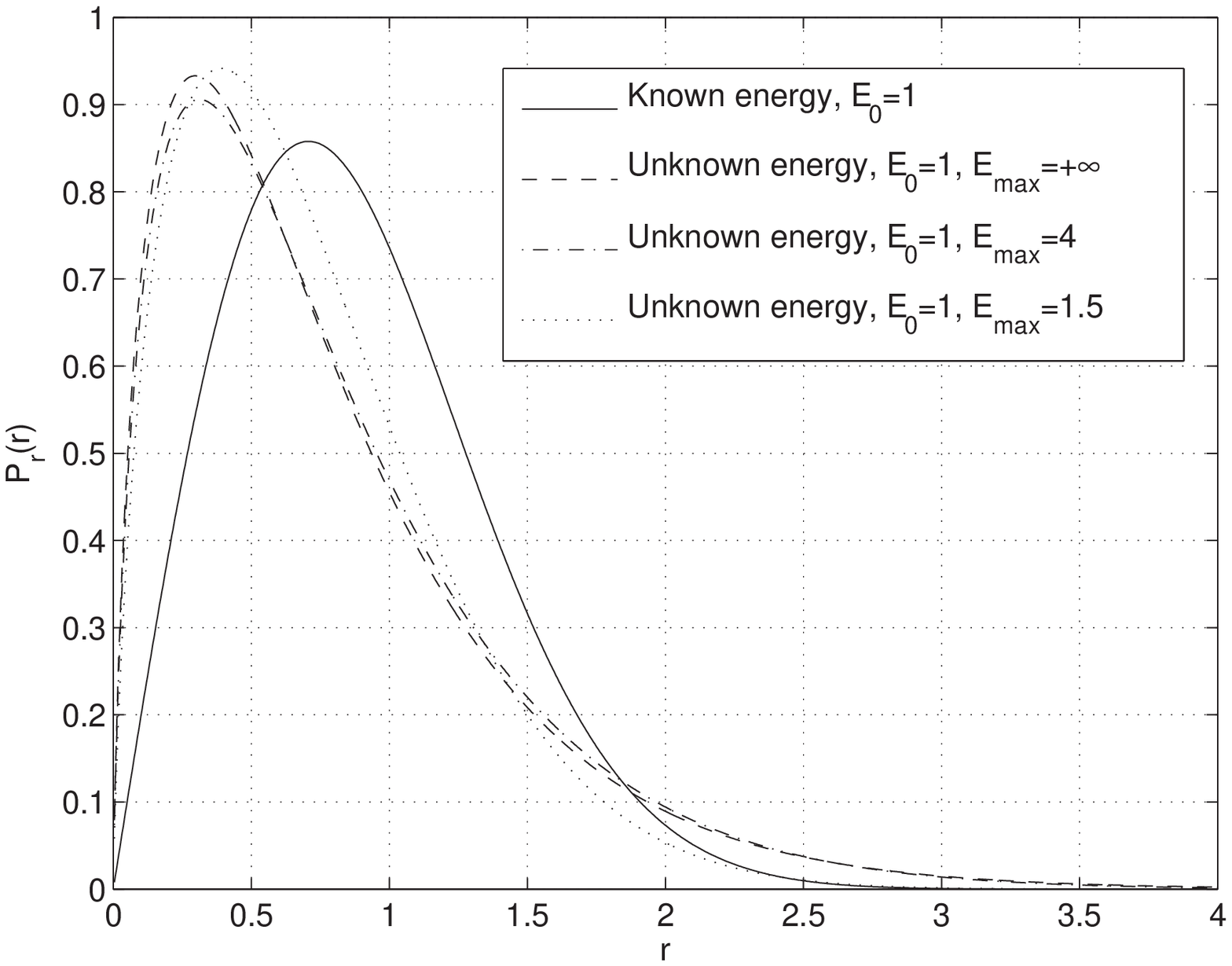,width=7.9cm}\label{siso_Pr}}
\subfigure[CDF of instantaneous mutual information $I(r)$]{\epsfig{figure=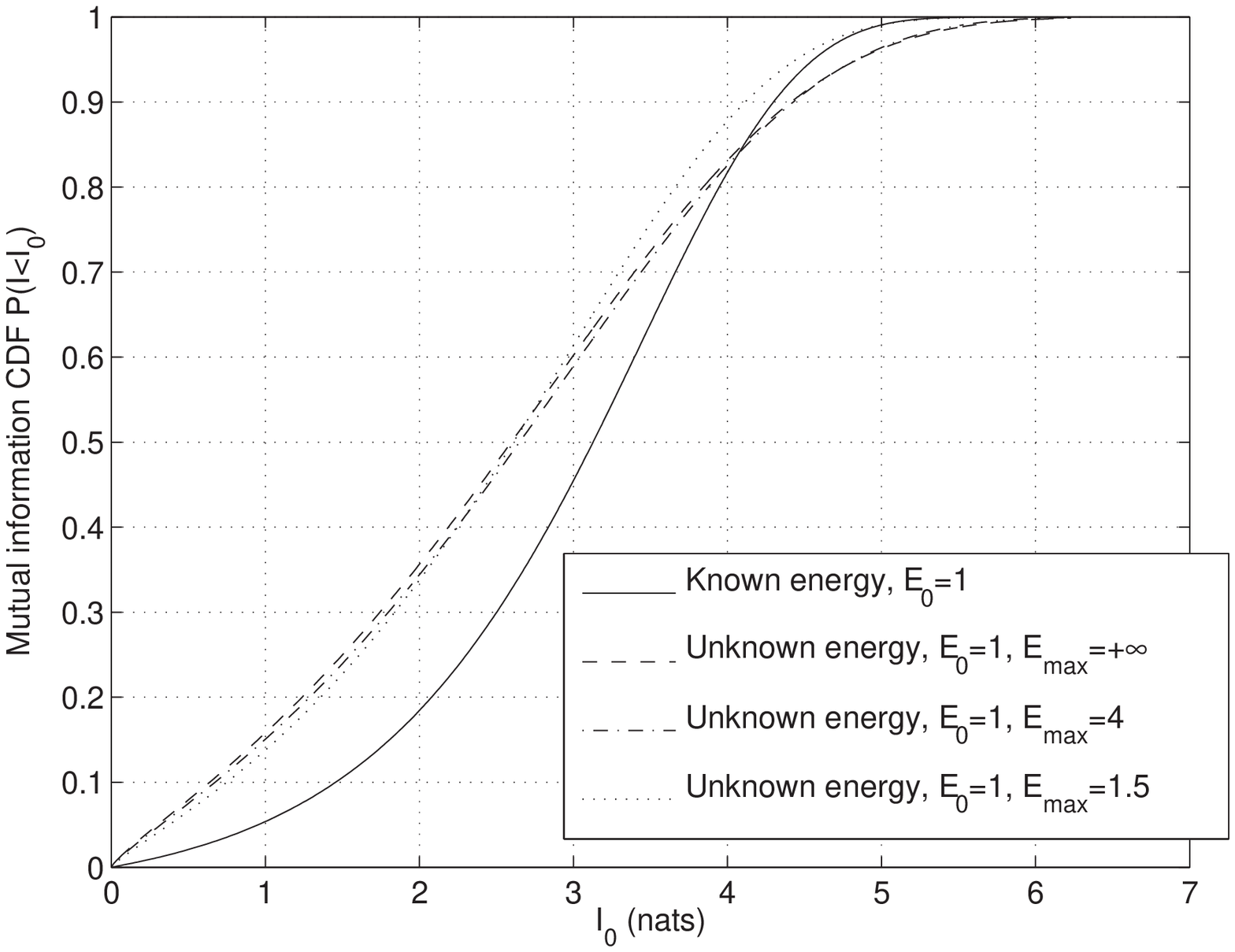,width=7.9cm}\label{siso_cdf}}
\caption{Amplitude and mutual information distributions of the proposed SISO channel models.}
\label{scalar_example_fig}
\end{figure}

In order to illustrate the difference between the two situations presented so far, let us investigate the Single-Input Single-Output (SISO) case $\Nt=\Nr=1$, where the channel is represented by a single complex scalar $h$.
Furthermore, since the distribution is circularly symmetric, it is more convenient to consider the distribution of $r\triangleq |h|$. After the change of variables $h\triangleq r (\cos \theta+i \sin \theta )$, and marginalization over $\theta$, eq.~(\ref{maxent_Giid}) becomes
\begin{equation}
  P_r(r) = \frac{2r}{E_0} \exp\left(-\frac{r^2}{E_0}\right), \label{scalar_eknown}
\end{equation}
whereas eq.~(\ref{unknown_E}) yields
\begin{equation}
  P_r(r) = \int_0^{E_{max}} \frac{\beta}{\exp(\beta E_{max})-1} \frac{2r}{E} \exp\left(\beta E-\frac{r^2}{E}\right) \d E. \label{scalar_eunknown}
\end{equation}
Note that the integral always exists since $\beta<0$. 
Figure~\ref{siso_Pr} depicts the probability density functions (PDFs) of $r$ under the known energy constraint (eq.~(\ref{scalar_eknown}), with $E_0=1$), and the known energy distribution constraint (eq.~(\ref{scalar_eunknown}) is computed numerically, for $E_{max}=1.5, 4$ and $+\infty$, taking $E_0=1$).
Figure~\ref{siso_cdf} depicts the cumulative density function (CDF) of the corresponding instantaneous mutual information $I(r)\triangleq \log(1+\rho r^2)$, for signal-to-noise ratio $\rho = 15 \ \mathrm{dB}$.
The lowest range of the CDF is of particular interest for wireless communications since it represents the probability of a channel outage for a given transmission rate. The curves clearly show that the models corresponding to the unknown energy have a lower outage capacity that the Gaussian channel model.

\section{Spatial correlation models}
\label{section_spatialcorrelation}
In this section, we shall incorporate several states of knowledge about the spatial correlation characteristics of the channel in the framework of maximum entropy modeling. We first study the case where the correlation matrix is deterministic, and subsequently extend the result to an unknown covariance matrix.

\subsection{Deterministic knowledge of the correlation matrix}

In this section, we establish the maximum entropy distribution of $\mathbf{H}$ under the assumption that the covariance matrix $\mathbf{Q} \triangleq \int_{\mathbb{C}^N} \v{h}\v{h}^H P_{\mathbf{H}|\mathbf{Q}}(\mathbf{H}) \d\mathbf{H}$ is known, where $\mathbf{Q}$ is a $N\times N$ complex Hermitian matrix.
Each component of the covariance constraint represents an independent linear constraint of the form 
\begin{equation}
  \int_{\mathbb{C}^N} h_{a} h_{b}^* P_{\mathbf{H}|\mathbf{Q}}(\mathbf{H}) \d\mathbf{H} = q_{a,b}
\end{equation}
for $(a,b)\in [1,\ldots,N]^2$. Note that this constraint makes any previous energy constraint redundant since $\int_{\mathbb{C}^N} ||\mathbf{H}||_{F}^2 P_{\mathbf{H}|\mathbf{Q}}(\mathbf{H}) \d\mathbf{H} =  \tr(\mathbf{Q})$.
Proceeding along the lines of the method exposed previously, we introduce $N^2$ Lagrange coefficients $\alpha_{a,b}$, and maximize
\begin{eqnarray}
 L(P_{\mathbf{H}|\mathbf{Q}}) = \int_{\mathbb{C}^N} -\log(P_{\mathbf{H}|\mathbf{Q}}(\mathbf{H})) P_{\mathbf{H}|\mathbf{Q}}(\mathbf{H}) \d\mathbf{H} &+& \beta \left[1- \int_{\mathbb{C}^N} P_{\mathbf{H}|\mathbf{Q}}(\mathbf{H}) \d\mathbf{H}\right] \nonumber \\
 &+&\sum_{\substack{a\in [1,\ldots,N]\\b\in [1,\ldots,N]}} \alpha_{a,b} \left[ \int_{\mathbb{C}^N} h_{a} h_{b}^* P_{\mathbf{H}|\mathbf{Q}}(\mathbf{H}) \d\mathbf{H} - q_{a,b}\right].
\end{eqnarray}
Denoting $\mathbf{A}=[\alpha_{a,b}]_{(a,b)\in [1,\ldots,N]^2}$ the $N\times N$ matrix of the Lagrange multipliers, the derivative is
\begin{equation}
  \frac{\delta L(P_{\mathbf{H}|\mathbf{Q}})}{\delta P_{\mathbf{H}|\mathbf{Q}}} = -\log(P_{\mathbf{H}|\mathbf{Q}}(\mathbf{H}))-1 -\beta -\v{h}^T \mathbf{A}\v{h}^*=0.
  \label{derivativePQ}
\end{equation}
Therefore, $P_{\mathbf{H}|\mathbf{Q}}(\mathbf{H}) = \exp\left( -(\beta+1)  -\v{h}^T \mathbf{A} \v{h}^* \right)$, or, after elimination of the Lagrange coefficients through proper normalization, 
\begin{equation}
  P_{\mathbf{H}|\mathbf{Q}}(\mathbf{H},\mathbf{Q}) = \frac{1}{ \det(\pi\mathbf{Q})}\exp\left(-( \v{h}^H \mathbf{Q}^{-1} \v{h} ) \right) \label{maxent_G_corr}.
\end{equation}
Again, the maximum entropy principle yields a Gaussian distribution, although of course its components are not independent anymore.

\subsection{Knowledge of the existence of a correlation matrix}
\label{section_maxent_Q}
It was shown in Section~\ref{previous_results} that in the absence of information on space correlation, maximum entropy modeling yields i.i.d. coefficients for the channel matrix, and therefore an identity covariance matrix. We now consider the case where covariance is known to be a parameter of interest, but is not known deterministically.
Again, we will proceed in two steps, first seeking a probability distribution function for the covariance matrix $\mathbf{Q}$, and then marginalizing the channel distribution over $\mathbf{Q}$.\\

\subsubsection{Correlation matrix PDF}
\label{sectionmaxentQ}

Let us first establish the distribution of $\mathbf{Q}$, under the energy constraint $\int \tr(\mathbf{Q}) P_{\mathbf{Q}}(\mathbf{Q}) \d\mathbf{Q}=N E_0$, by maximizing the functional
\begin{equation}
  L(P_{\mathbf{Q}}) = \int_{\mathcal{S}} -\log(P_{\mathbf{Q}}(\mathbf{Q})) P_{\mathbf{Q}}(\mathbf{Q}) \d\mathbf{Q} + \beta \left[ \int_{\mathcal{S}}  P_{\mathbf{Q}}(\mathbf{Q}) \d\mathbf{Q} -1 \right] + \gamma \left[ \int_{\mathcal{S}}  \tr(\mathbf{Q}) P_{\mathbf{Q}}(\mathbf{Q}) \d\mathbf{Q} - N E_0 \right].   \label{maxentQ}
\end{equation}

Due to their structure, covariance matrices are restricted to the space $\mathcal{S}$ of $N\times N$ positive semidefinite complex matrices. Therefore, let us perform the variable change to the eigenvalues/eigenvectors space. Specifically, let us denote $\Lambda \triangleq \diag(\lambda_1 \ldots \lambda_N)$ the diagonal matrix containing the eigenvalues of $\mathbf{Q}$, and let $\mathbf{U}$ be the unitary matrix containing the eigenvectors, such that $\mathbf{Q} = \mathbf{U}\Lambda\mathbf{U}^H$.\\

We use the mapping between the space of complex $N\times N$ self-adjoint matrices (of which $\mathcal{S}$ is a subspace), and $\mathcal{U}(N)/T  \times \mathbb{R}_{\leq}^N$, where $\mathcal{U}(N)/T$ denotes the space of unitary $N\times N$ matrices with real, non-negative first row, and $\mathbb{R}_{\leq}^N$ is the space of real N-tuples with non-decreasing components (see \cite[Lemma 4.4.6]{Hiai_Petz_monograph}). The positive semidefinite property of the covariance matrices further restricts the components of $\Lambda$ to non-negative values, and therefore $\mathcal{S}$ maps into $\mathcal{U}(N)/T  \times {\mathbb{R}_{\leq}^+}^N$.\\

Let us now define function $F$ over $\mathcal{U}(N)/T \times {\mathbb{R}_{\leq}^+}^N$ as
\begin{equation}
  F(\mathbf{U,\Lambda}) = P_{\mathbf{Q}}(\mathbf{U}\Lambda\mathbf{U}^H), \quad \mathbf{U} \in \mathcal{U}(N)/T, \quad \Lambda \in {\mathbb{R}_{\leq}^+}^N.
\end{equation}
According to this mapping, eq.~(\ref{maxentQ}) becomes
\begin{eqnarray}
  L(F) &=& \int_{\mathcal{U}(N)/T \times \mathbb{R}_{\leq}^{+N}} -\log(F(\mathbf{U,\Lambda})) F(\mathbf{U,\Lambda})  K(\Lambda) \d\mathbf{U}\d{\Lambda} \nonumber \\
  &+& \beta \left[ \int_{\mathcal{U}(N)/T \times \mathbb{R}_{\leq}^{+N}}  F(\mathbf{U,\Lambda}) K(\Lambda) \d\mathbf{U}\d{\Lambda} -1 \right] \nonumber \\
  & +& \gamma \left[ \int_{\mathcal{U}(N)/T \times \mathbb{R}_{\leq}^{+N}} \left( \sum_{i=1}^N \lambda_i \right) F(\mathbf{U,\Lambda}) K(\Lambda) \d\mathbf{U}\d{\Lambda}- N E_0 \right],   \label{maxentULambda}
\end{eqnarray}
where we introduced the corresponding Jacobian $K(\Lambda) \triangleq \frac{(2\pi)^{N(N-1)/2}}{\prod_{j=1}^{N-1} j!} \prod_{i<j} (\lambda_i-\lambda_j)^2$, and used \linebreak $\tr(\mathbf{Q})= \tr(\Lambda) = \sum_{i=1}^N \lambda_i$. Maximizing the entropy of the distribution $P_{\mathbf{Q}}$ by taking $\frac{\delta L(F)}{\delta F} = 0$ yields
\begin{equation}
  -K(\Lambda) -K(\Lambda)\log(F(\mathbf{U,\Lambda})) + \beta K(\Lambda) +\gamma \left(\sum_{i=1}^N \lambda_i \right)K(\Lambda) =0.
\end{equation}
Since $K(\Lambda)\neq 0$ except on a set of measure zero, this is equivalent to
\begin{equation}
  F(\mathbf{U,\Lambda}) = \exp\left( \beta-1+\gamma \sum_{i=1}^N \lambda_i\right). \label{P_U_Lambda}
\end{equation}
Note that the distribution $F(\mathbf{U,\Lambda})K(\Lambda)$ does not explicitly depend on $\mathbf{U}$. This indicates that $\mathbf{U}$ is uniformly distributed, with constant density $P_{\mathbf{U}}=(2\pi)^N$ over $\mathcal{U}(N)/T$. 
Therefore, the joint density can be factored as $F(\mathbf{U,\Lambda})K(\Lambda) = P_{\mathbf{U}} P_{\mathbf{\Lambda}}(\mathbf{\Lambda})$, where the distribution of the eigenvalues over ${\mathbb{R}_{\leq}^+}^N$ is
\begin{equation}
P_{\Lambda}(\Lambda) = \frac{\e^{\beta-1}}{P_{\mathbf{U}}} \exp\left(\gamma \sum_{i=1\ldots N} \lambda_i\right) \frac{(2\pi)^{N(N-1)/2}}{\prod_{j=1}^{N-1} j!} \prod_{i<j} (\lambda_i-\lambda_j)^2. \label{orderpdflambda}
\end{equation}

At this point, it is worth noting that the form of eq.~(\ref{orderpdflambda}) indicates that the order of the eigenvalues is immaterial. 
In order to see this, consider a pair of eigenvalues $(\lambda_i,\lambda_j)$, with $i<j$ and $\lambda_i\leq \lambda_j$, and the change of variables $(x,y)=\left(\frac{\lambda_i+\lambda_j}{\sqrt{2}},\frac{-\lambda_i+\lambda_j}{\sqrt{2}} \right)$.
For any function $f(\lambda_i,\lambda_j)$, 
\begin{equation}
  \int_{0\leq \lambda_i \leq \lambda_j \leq +\infty} f(\lambda_i,\lambda_j) \d \lambda_i \d \lambda_j = \int_{x=0}^{+\infty} \int_{y=0}^{x} f(x,y) \d y \d x,
\end{equation}
whereas for the the non-restricted integral 
\begin{equation}
  \int_{(\lambda_i,\lambda_j)\in \mathbb{R}^{+2}} f(\lambda_i,\lambda_j) \d \lambda_i \d \lambda_j = \int_{x=0}^{+\infty} \int_{y=-x}^{x} f(x,y) \d y \d x.
\end{equation}
Note that for every function $f$ s.t. $f(x,y)=f(x,-y)$,
\begin{equation}
 \int_{y=-x}^{x} f(x,y) \d y = 2 \int_{y=0}^{x} f(x,y) \d y,
\end{equation}
and therefore
\begin{equation}
  \int_{(\lambda_i,\lambda_j)\in \mathbb{R}^{+2}} f(\lambda_i,\lambda_j) \d \lambda_i \d \lambda_j = 2  \int_{0\leq \lambda_i \leq \lambda_j \leq +\infty} f(\lambda_i,\lambda_j) \d \lambda_i \d \lambda_j.
\end{equation}
Since the probability distribution $P_{\Lambda}(\Lambda)$ in (\ref{orderpdflambda}) obviously verifies the property $f(x,y)=f(x,-y)$ in the rotated space for any $0\leq i,j \leq N$, this reasoning (generalized to any permutation of the ordered eigenvalues) applies to $P_{\Lambda}(\Lambda)$. Therefore, for the sake of simplicity, we will now work with the PDF $P'_{\Lambda}(\Lambda)$ of the joint distribution of the \emph{unordered} eigenvalues, defined over ${\mathbb{R}^+}^N$.
Note that its restriction to the set of the ordered eigenvalues is proportional to $P_{\Lambda}(\Lambda)$. More precisely, 
\begin{equation}
 \forall \Lambda \in {\mathbb{R}^+}^N, \quad P'_{\Lambda}(\Lambda) = \frac{1}{N!} P_{\Lambda}(\lambda_{s(1)},\ldots,\lambda_{s(N)}) \end{equation}
where $s$ is any permutation of $\{1\ldots N\}$ such that $\lambda_{s(1)} \leq \lambda_{s(2)} \leq \ldots\leq  \lambda_{s(N)}$, and the coefficient $1/N!$ comes from the number of permutations of the $N$ eigenvalues.
Since $P_{\Lambda}(\lambda_{s(1)},\ldots,\lambda_{s(N)})=P_{\Lambda}(\lambda_1,\ldots,\lambda_N)$,
this yields
\begin{equation}
 P'_{\Lambda}(\Lambda) = C \exp\left(\gamma \sum_{i=1\ldots N} \lambda_i\right)  \prod_{i<j} (\lambda_i-\lambda_j)^2 \label{unorderedpdflambda},
\end{equation}
where the value of $C=\frac{\e^{\beta-1}}{P_{\mathbf{U}}} \frac{(2\pi)^{N(N-1)/2}}{ N!\prod_{j=1}^{N-1} j!}$ can be determined by 
solving the normalization equation for the probability distribution $P'_{\Lambda}$:
\begin{eqnarray}
  1 = \int_{{\mathbb{R}^+}^N} P'_{\Lambda}(\Lambda)\d\Lambda &=& C \int_{{\mathbb{R}^+}^N}    \prod_{i=1}^N \e^{\gamma \lambda_i} \prod_{i<j} (\lambda_i-\lambda_j)^2 \d\Lambda \\
  &=& C \left( -\frac{1}{\gamma}\right)^{N^2} \int_{{\mathbb{R}^+}^N}  \prod_{i=1}^N \e^{-x_i} \prod_{i<j} (x_i-x_j)^2 \d x_1 \ldots \d x_N\\
  &=& C \left( -\frac{1}{\gamma}\right)^{N^2} \prod_{i=0}^{N-1} \frac{\Gamma(i+2) \Gamma(i+1)}{\Gamma(2)} = C \left( -\frac{1}{\gamma}\right)^{N^2} \prod_{n=1}^N  n! (n-1)!,
\end{eqnarray}
where we used the change of variables $x_i=-\gamma \lambda_i$ and the Selberg integral (see \cite[eq.~(17.6.5)]{Mehta_random_matrices}). This yields $C=(-\gamma)^{N^2} \prod_{n=1}^N [n! (n-1)!]^{-1}$. Furthermore, $\frac{\d \log(C)}{\d (-\gamma)}=\frac{N^2}{-\gamma}=N E_0$, and we finally obtain the final expression of the eigenvalue distribution
\begin{equation}
  P'_{\Lambda}(\Lambda) = \left(\frac{N}{E_0} \right)^{N^2} \prod_{n=1}^N \frac{1}{n! (n-1)!} \exp\left(-\frac{N}{E_0} \sum_{i=1\ldots N} \lambda_i\right) \prod_{i<j} (\lambda_i-\lambda_j)^2 \label{finalpdflambda}.\\
\end{equation}

In order to obtain the final distribution of $\mathbf{Q}$, let us first note that since the order of the eigenvalues has been shown to be immaterial, the restriction of $\mathbf{U}$ to $\mathcal{U}(N)/T$ is not necessary, and $\mathbf{Q}$ is distributed as $\mathbf{U}\Lambda\mathbf{U}^H$, where the distribution of $\Lambda$ is given by eq.~(\ref{finalpdflambda}) and $\mathbf{U}$ is Haar distributed (uniform on $\mathcal{U}(N)$).  Furthermore, note that eq.~(\ref{finalpdflambda}) is a particular case of the density of the eigenvalues of a complex Wishart matrix \cite{tulino_random_matrices,Edelman_PhD}. We recall that the complex $N \times N$ Wishart matrix with $K$ degrees of freedom and covariance $\mathbf{\Sigma}$  (denoted by $\tilde{\mathcal{W}}_N(K,\mathbf{\Sigma})$) is the matrix $\mathbf{A}=\mathbf{B}\mathbf{B}^H$ where $\mathbf{B}$ is a $N\times K$ matrix whose columns are complex circularly-symmetric independent Gaussian vectors with covariance $\mathbf{\Sigma}$.
Indeed, eq.~(\ref{finalpdflambda}) describes the unordered eigenvalue density of a $\tilde{\mathcal{W}}_N(N,\frac{E_0}{N}\Id_N)$ matrix. Taking into account the isotropic property of the distribution of $\mathbf{U}$, we can conclude that $\mathbf{Q}$ itself is also a $\tilde{\mathcal{W}}_N(N,\frac{E_0}{N}\Id_N)$ Wishart matrix. A similar result, with a slightly different constraint, was obtained by Adhikari in \cite{Adhikari_maxent_elastodynamics2006}, where it is shown that the entropy-maximizing distribution of a positive definite matrix with known mean $\mathbf{G}$ follows a Wishart distribution with $N+1$ degrees of freedom, more precisely the $\tilde{\mathcal{W}}_N(N+1,\frac{\mathbf{G}}{N+1})$ distribution.\\

The isotropic property of the obtained Wishart distribution (due to the fact that $\mathbf{U}$ is Haar distributed, \emph{i.e.} there is not privileged direction for the eigenvalues of the covariance matrix $\mathbf{Q}$), is a consequence of the fact that no spatial constraints were imposed on the correlation. The energy constraint (imposed through the trace) only affects the distribution of the eigenvalues of $\mathbf{Q}$.
Note also that the generation  for simulation purposes of $\mathbf{Q}$ according to the Wishart distribution obtained above is easy, since it can be obtained as $\mathbf{Q}=\frac{E_0}{N}\mathbf{B}\mathbf{B}^H$, where $\mathbf{B}$ is a $N \times N$ matrix with i.i.d. complex circularly-symmetric Gaussian coefficients of unit variance.

\subsubsection{Application to the Kronecker channel model}
We highlight the fact that the result of Section~\ref{section_maxent_Q} is directly applicable to the case where the channel correlation is known to be separable between transmitter and receiver. In this case \cite{Chuah_Kronecker_model}, the full correlation matrix $\mathbf{Q}$ is known to be the Kronecker product of the transmit and receive correlation matrices, \emph{i.e.} $\mathbf{Q}=\mathbf{Q}_T \otimes \mathbf{Q}_R$, where $\mathbf{Q}_T$ and $\mathbf{Q}_R$ are respectively the transmit and receive correlation matrices. This channel model is therefore denoted by ``Kronecker model'', see \cite{Oestges_Kronecker_validity_vtcsp06} for an overview of its applicability.
The stochastic nature of $\mathbf{Q}_T$ and $\mathbf{Q}_R$ is barely mentioned in the literature, since the correlation matrices are usually assumed to be measurable quantities associated to a particular antenna array shape and propagation environment.
However, in situations where these are not known (for instance, if the array shape is not known at the time of the channel code design, or if the properties of the scattering environment can not be determined), but the Kronecker model is assumed to hold, our analysis suggests that the maximum entropy choice for the distribution of $\mathbf{Q}_T$ and $\mathbf{Q}_R$ is independent, complex Wishart distributions with respectively $\Nt$ and $\Nr$ degrees of freedom.

\subsubsection{Marginalization over $\mathbf{Q}$}
\label{section_marginalizeQ}

The complete distribution of the correlated channel can be obtained by marginalizing out $\mathbf{Q}$, using its distribution as established in Section \ref{sectionmaxentQ}.
The distribution of $\mathbf{H}$ is obtained through 
\begin{equation}
  P_{\mathbf{H}}(\mathbf{H}) = \int_{\mathcal{S}} P_{\mathbf{H}|\mathbf{Q}}(\mathbf{H},\mathbf{Q})  P_{\mathbf{Q}}(\mathbf{Q}) \d\mathbf{Q} = \int_{\mathcal{U}(N) \times{\mathbb{R}^+}^N}  P_{\mathbf{H}|\mathbf{Q}}(\mathbf{H},\mathbf{U},\Lambda)  P'_{\mathbf{\Lambda}}(\mathbf{\Lambda})  \d\mathbf{U}\d\Lambda \label{integralH_existingQ}
\end{equation}

Let us rewrite the conditional probability density of eq.~(\ref{maxent_G_corr}) as
\begin{equation}
  P_{\mathbf{H}|\mathbf{Q}}(\v{h},\mathbf{U}, \Lambda) = \frac{1}{\pi^N \det(\Lambda)} \e^{-\v{h}^H \mathbf{U}\Lambda^{-1}\mathbf{U}^H \v{h} } = \frac{1}{\pi^N \det(\Lambda)} \e^{ -\tr( \v{h}\v{h}^H \mathbf{U}\Lambda^{-1}\mathbf{U}^H   )}. \label{conditional_H_Q}
\end{equation}
Using this expression in (\ref{integralH_existingQ}), we obtain
\begin{equation}
P_{\mathbf{H}}(\mathbf{H}) = \frac{1}{\pi^N} \int_{{\mathbb{R}^+}^N} \int_{\mathcal{U}(N)}   \e^{ -\tr( \v{h}\v{h}^H\mathbf{U}\Lambda^{-1}\mathbf{U}^H )}   \d\mathbf{U} \, \det(\Lambda)^{-1} P'_{\mathbf{\Lambda}}(\mathbf{\Lambda}) \d\Lambda. \label{integralH_intermed1}
\end{equation}

Following the notations of \cite{Capacity_and_character_expansion_IT}, let $\det(f(i,j))$ denote the determinant of a matrix with the $(i,j)$-th element given by an arbitrary function $f(i,j)$.
Also, let $\Delta(\mathbf{X})$ denote the Vandermonde determinant of the eigenvalues $x_i$ of the matrix $\mathbf{X}$
\begin{equation}
\Delta(\mathbf{X}) = \det(x_i^{j-1}) = \prod_{i>j} (x_i - x_j). \label{def_Vandermonde_determinant}
\end{equation}
Using these notations, let us recall the Harish-Chandra-Itzykson-Zuber (HCIZ) integral \cite{Balantekin_Un_character_expansion}
\begin{equation}
  \int_{\mathcal{U}(N)} \exp( \kappa \tr(\mathbf{AUBU}^H)) \d \mathbf{U} = \left( \prod_{n=1}^{N-1} n! \right) \kappa^{N(N-1)/2} \frac{\det\left( e^{-A_i B_j} \right)}{\Delta(\mathbf{A}) \Delta(\mathbf{B})}, \label{HCIZ}
\end{equation}
where $\mathbf{A}$ and $\mathbf{B}$ are any hermitian matrices with respective eigenvalues $A_1,\ldots,A_N$ and $B_1,\ldots,B_N$. Let us explicit the Haar integral in (\ref{integralH_intermed1}) using the Harish-Chandra-Itzykson-Zuber result by identifying $\mathbf{A}=\v{h}\v{h}^H$ and $\mathbf{B}=\Lambda^{-1}$. Note however that we can not directly apply the result in (\ref{HCIZ}) since $\mathbf{A}$ is rank one, and therefore $\det \mathbf{A}=0$.
This can be resolved by taking the limit of all other eigenvalues to zero one by one, and applying the l'Hospital rule.
Therefore, let $\mathbf{A}$ be an Hermitian matrix which has its $N$th eigenvalue $A_N$ equal to $\v{h}^H\v{h}$, and the others $A_1,\ldots, A_{N-1}$ are arbitrary, positive values that will eventually be set to 0. Letting
\begin{equation}
  I(\mathbf{H},A_1,\ldots, A_{N-1})=\frac{1}{\pi^N} \int_{{\mathbb{R}^+}^N} \int_{\mathcal{U}(N)}   \e^{ -\tr( \mathbf{A}\mathbf{U}\Lambda^{-1}\mathbf{U}^H )}  P_{\mathbf{U}}   \d\mathbf{U} \, \det(\Lambda)^{-1} P'_{\mathbf{\Lambda}}(\mathbf{\Lambda}) \d\Lambda,
\end{equation}
$P_{\mathbf{H}}(\mathbf{H})$ can be determined as the limit distribution when the first $N-1$ eigenvalues of $\mathbf{A}$ go to zero:
\begin{equation}
  P_{\mathbf{H}}(\mathbf{H}) = \lim_{A_1,\ldots,A_{N-1} \rightarrow 0} I(\mathbf{H},A_1,\ldots, A_{N-1}).
\end{equation}

Applying the HCIZ to integrate over $\mathbf{U}$ yields
\begin{eqnarray}
  I(\mathbf{H},A_1,\ldots, A_{N-1}) &=&\frac{(-1)^{\frac{N(N-1)}{2}}}{\pi^N}\left(\prod_{n=1}^{N-1} n!\right) \int_{{\mathbb{R}^+}^N} \frac{\det\left( e^{-A_i/\lambda_j} \right)}{\Delta(\mathbf{A})\Delta(\Lambda^{-1})} \det(\Lambda)^{-1} P'_{\mathbf{\Lambda}}(\mathbf{\Lambda}) \d\Lambda \\
  &=& \frac{1}{\pi^N} \left(\prod_{n=1}^{N-1} n!\right)  \int_{{\mathbb{R}^+}^N}   \frac{\det\left( e^{-A_i/\lambda_j} \right) \det(\Lambda)^{N-2}}{\Delta(\mathbf{A}) \Delta(\Lambda)}  P'_{\mathbf{\Lambda}}(\mathbf{\Lambda}) \d\Lambda \\
  &=& \frac{C}{\pi^N} \left(\prod_{n=1}^{N-1} n!\right)  \int_{{\mathbb{R}^+}^N}   \frac{\det\left( e^{-A_i/\lambda_j} \right) \det(\Lambda)^{N-2}\Delta(\Lambda)}{\Delta(\mathbf{A})} \e^{-\frac{N}{E_0}\tr(\Lambda)} \d\Lambda,
\end{eqnarray}
where we used the identity $\Delta(\Lambda^{-1}) = \det(\frac{1}{\lambda_i}^{j-1}) = (-1)^{N(N+3)/2} \frac{\Delta(\Lambda)}{\det(\Lambda)^{N-1}}$.

Then, let us decompose the determinant product using the expansion formula: for an arbitrary $N\times N$ matrix $\mathbf{X}=(X_{i,j})$,
\begin{equation}
  \det(\mathbf{X}) = \sum_{\mathbf{a}\in \mathcal{P}_N} (-1)^{\mathbf{a}} \prod_{n=1}^N X_{n,a_n} = \frac{1}{N!} \sum_{\mathbf{a},\mathbf{b}\in \mathcal{P}_N} (-1)^{\mathbf{a}+\mathbf{b}} \prod_{n=1}^N X_{a_n,b_n}, \label{determinant_expansions}
\end{equation}
where $\mathbf{a}=[a_1,\ldots,a_N]$, $\mathcal{P}_N$ denotes the set of all permutations of $[1,\ldots,N]$, and $(-1)^{\mathbf{a}}$ is the sign of the permutation.
Using the first form of the expansion, we obtain
\begin{eqnarray}
  \Delta(\Lambda)\det\left( e^{-A_i/\lambda_j} \right) &= & \det(\lambda_i^{j-1})\det( e^{-A_j/\lambda_i} ) \label{playwithdets}\\
  & = &  \left( \sum_{\mathbf{a}\in \mathcal{P}_N} (-1)^{\mathbf{a}} \prod_{n=1}^N \lambda_n^{a_n-1} \right) \left( \sum_{\mathbf{b}\in \mathcal{P}_N} (-1)^{\mathbf{b}} \prod_{m=1}^N \e^{-A_{b_m}/{\lambda_m}} \right) \\
  & = &  \sum_{\mathbf{a,b}\in \mathcal{P}_N^2} (-1)^{\mathbf{a}+\mathbf{b}} \prod_{n=1}^N \lambda_n^{a_n-1} \e^{-A_{b_n}/{\lambda_n}}.
\end{eqnarray}
Note that in (\ref{playwithdets}) we used the invariance of the second determinant by transposition in order to simplify subsequent derivations. Therefore,
\begin{eqnarray}
  I(\mathbf{H},A_1,\ldots, A_{N-1}) &=& \frac{C}{\pi^N \Delta(\mathbf{A})} \left(\prod_{n=1}^{N-1} n!\right) \int_{{\mathbb{R}^+}^N}   \sum_{\mathbf{a,b}\in \mathcal{P}_N} (-1)^{\mathbf{a}+\mathbf{b}} \prod_{n=1}^N \lambda_n^{N+a_n-3} \e^{-A_{b_n}/{\lambda_n}} \e^{-\frac{N}{E_0} \lambda_n}  \d\Lambda \\
  &=& \frac{C}{\pi^N \Delta(\mathbf{A})} \left(\prod_{n=1}^{N-1} n!\right)  \sum_{\mathbf{a,b}\in \mathcal{P}_N} (-1)^{\mathbf{a}+\mathbf{b}} \prod_{n=1}^N \int_{\mathbb{R}^+} \lambda_n^{N+a_n-3} \e^{-A_{b_n}/{\lambda_n}} \e^{-\frac{N}{E_0} \lambda_n} \d \lambda_n \\
  &=& \frac{C \, N!}{\pi^N } \left(\prod_{n=1}^{N-1} n!\right)  \frac{\det[f_i(A_j)]}{\Delta(\mathbf{A})},
\end{eqnarray}
where we let $f_i(x)= \int_{\mathbb{R}^+} t^{N+i-3} \e^{-x/t} \e^{-\frac{N}{E_0} t} \d t$, and recognize the second form of the determinant expansion  (eq.~(\ref{determinant_expansions})).
In order to obtain the limit as $A_1,\ldots A_{N-1}$ go to zero, we use a result from \cite[Appendix III]{Capacity_and_character_expansion_IT}, which states that the limit of the ratio $\frac{\det(f_i(x_j))}{\Delta(\mathbf{X})}$ of the singular determinants as the first $p$ eigenvalues go to $x_0$ is
\begin{equation}
  \lim_{x_1,x_2,\ldots,x_p \rightarrow x_0} \frac{\det(f_i(x_j))}{\Delta(\mathbf{X})} = \frac{\det\left[f_i(x_0);f_i'(x_0);\ldots;f_i^{(p-1)}(x_0);f_i(x_{p+1});\ldots; f_i(x_N) \right]}{\Delta(x_{p+1},\ldots,x_N) \prod_{i=p+1}^N (x_i-x_0)^p \prod_{j=1}^{p-1} j!},
\end{equation}
where the first $p$ columns in the right-hand side determinant represent the successive derivatives of the function $f$, and the rows correspond to different values of $i=1,\ldots, N$.
Applying this -- with $p=N-1$ and $x_0=0$ since $\mathbf{A}$ has only one non-zero eigenvalue -- yields
\begin{eqnarray}
  P_{\mathbf{H}}(\mathbf{H}) &=& \lim_{A_1,A_2,\ldots,A_{N-1} \rightarrow 0}  I(\mathbf{H},A_1,\ldots, A_{N-1}) \\
  &=& \frac{(-\gamma)^{N^2}}{\pi^N x_N^{N-1}} \prod_{n=1}^{N-1} \left[ n! (n-1)! \right]^{-1}  \det\left[f_i(0);f_i'(0);\ldots;f_i^{(N-2)}(0); f_i(x_N) \right]. \label{P_h_determinant}
\end{eqnarray}

At this point, it becomes obvious from (\ref{P_h_determinant}) that the probability of $\mathbf{H}$ depends only on its norm (recall that $x_N=\v{h}^H\v{h}$ by definition of $\mathbf{A}$). The distribution of $\v{h}$ is isotropic, and is completely determined by the probability density function $P_{x}(x)$ of having $\v{h}$ s.t. $\v{h}^H\v{h}=x$.\\

Therefore, for a given $x$, $\v{h}$ is uniformly distributed over $\mathbb{S}^{N-1}(x)= \left\{\v{h} \, \mathrm{s.t.} \, \v{h}^H\v{h} = x \right\}$, the zero-centered complex hypersphere of radius $x$. Its volume is $V_N(x)=\frac{\pi^N x^N}{N!}$, and its surface is $S_N(x)=\frac{\d V_N(x)}{\d x}=\frac{\pi^N x^{N-1}}{(N-1)!}$.
Therefore, we can write the probability density function of $x_N$ as
\begin{equation}
 P_{x}(x)= \int_{\mathbb{S}^{N-1}(x)} P_{\mathbf{H}}(\v{h}) \d \v{h} = \frac{(-\gamma)^{N^2}}{(N-1)!} \prod_{n=1}^{N-1} \left[ n! (n-1)! \right]^{-1}  \det\left[f_i(0);f_i'(0);\ldots;f_i^{(N-2)}(0); f_i(x) \right].
\end{equation}

In order to simplify the expression of the successive derivatives of $f_i$, it is useful to identify the Bessel $K$-function \cite[Section 8.432]{Gradshteyn_Ryzhik}, and to replace it by its infinite sum expansion \cite[Section 8.446]{Gradshteyn_Ryzhik} 
\begin{eqnarray}
  f_i(x) &=& 2\left(\sqrt{ \frac{x}{-\gamma}} \right)^{i+N-2} K_{i+N-2}(2\sqrt{-\gamma x})\\
  &=& (-\gamma)^{-i-N+2} \left[ \sum_{k=0}^{i+N-3} (-1)^k \frac{(i+N-3-k)!}{k!} (-\gamma x)^k + \right. \nonumber \\  && \left. (-1)^{i+N-1} \sum_{k=0}^{+\infty} \frac{(-\gamma x)^{i+N-2+k}}{k! (i+N-2+k)!} \left( \ln(-\gamma x)-\psi(k+1)-\psi(i+N-1+k) \right) \right].
\end{eqnarray}
Note that there is only on term in the sum with a non-zero $p$th derivative at 0. Therefore, the $p$th derivative of $f_i$ at 0 is simply (for $0\leq p \leq N-2$)
\begin{equation}
  f_i^{(p)}(0) = (-1)^{-i-N} \gamma^{p-i-N+2} (i+N-3-p)!
\end{equation}
Let us bring the last column to become the first, and expand the resulting determinant along its first column:
\begin{eqnarray}
 \det\left[f_i^{(0)}(0);\ldots;f_i^{(N-2)}(0); f_i(x) \right] &=& (-1)^{N-1} \det\left[f_i(x); f_i^{(0)}(0);\ldots;f_i^{(N-2)}(0) \right] \\
 & =& (-1)^{N-1} \sum_{n=1}^{N} (-1)^{1+n} f_n(x) \det\left[\tilde{f}_{i,n}^{(0)}(0);\ldots;\tilde{f}_{i,n}^{(N-2)}(0) \right]
\end{eqnarray}
where $\tilde{f}_{i,n}^{(p)}(0)$ is the $N-1$ dimensional column obtained by removing the $n$th element from $f_i^{(p)}(0)$.
Factoring the $(-1)^p \gamma^{p-i-N+2}$ in the expression of $f_i^{(p)}(0)$ out of the determinant  yields
\begin{equation}
  \det\left[\tilde{f}_{i,n}^{(0)}(0);\ldots;\tilde{f}_{i,n}^{(N-2)}(0) \right] = (-1)^{n-N(N+1)/2} \gamma^{n-N^2+N-1} \det(\mathbf{g}^{(n)})
\end{equation}
where the $N-1$ dimensional matrix $\mathbf{g}^{(n)}$ has the elements
\begin{equation}
  \mathbf{g}_{l,k}^{(n)} = \Gamma(q_l^{(n)}+N-k-1)
\end{equation}
where
\begin{equation}
  q_l^{(n)}= \left\{ \begin{array}{ll}
   l, & l \leq n-1, \\
   l+1, & l \geq n.
  \end{array} \right.
\end{equation}
Using the fact that $\Gamma(q_l^{(n)}+i)= q_l^{(n)} \Gamma(q_l^{(n)}+i-1)+ (i-1) \Gamma(q_l^{(n)}+i-1)$, note that the $k$th column of $\mathbf{g}^{(n)}$ is
\begin{equation}
  \mathbf{g}_{l,k}^{(n)}=q_l^{(n)} \Gamma(q_l^{(n)}+N-k-2)+ (N-k-2) \mathbf{g}_{l,k+1}^{(n)}.
\end{equation}
Since the second term is proportional to the $(k+1)$th column, it can be omitted without changing the value of the determinant. Applying this property to the first $N-2$ pairs of consecutive columns, and repeating this process again to the first $N-2,\ldots,1$ pairs of columns, we obtain
\begin{eqnarray}
\det(\mathbf{g}^{(n)})&=& \det\left( \Gamma(q_l^{(n)}+N-2) ; \ldots ; \Gamma(q_l^{(n)}+2) ;\Gamma(q_l^{(n)}+1) ; \Gamma(q_l^{(n)})  \right)\\
&=& \det\left( q_l^{(n)} \Gamma(q_l^{(n)}+N-3) ; \ldots ; q_l^{(n)} \Gamma(q_l^{(n)}+1) ;q_l^{(n)} \Gamma(q_l^{(n)}) ; \Gamma(q_l^{(n)})  \right) \\
&=& \det\left( {q_l^{(n)}}^2 \Gamma(q_l^{(n)}+N-4) ; \ldots ; {q_l^{(n)}}^2 \Gamma(q_l^{(n)}) ;q_l^{(n)} \Gamma(q_l^{(n)}) ; \Gamma(q_l^{(n)})  \right) \\
&=& \det\left( {q_l^{(n)}}^{N-1-k} \Gamma(q_l^{(n)})  \right) \\
&=& \frac{\prod_{i=1}^N \Gamma(i)}{\Gamma(n)} \det\left({q_l^{(n)}}^{N-1-k} \right) \\
&=&  \frac{\prod_{i=1}^N \Gamma(i)}{\Gamma(n)} (-1)^{(N-1)(N-2)/2} \det\left({q_l^{(n)}}^{k-1} \right),
\end{eqnarray}
where the last two equalities are obtained respectively by factoring out the $\Gamma(q_l^{(n)})$ factors (common to all terms on the $l$th row) and inverting the order of the columns in order to get a proper Vandermonde structure.
Finally, the determinant can be computed using (\ref{def_Vandermonde_determinant}), as
\begin{eqnarray}
  \det\left({q_l^{(n)}}^{k-1} \right) &=& \prod_{1\leq j < i \leq N-1} \left( q_i^{(n)} - q_j^{(n)}\right) \\
  &=& \left(\prod_{1\leq j < i \leq n-1} (i-j)\right) \left(\prod_{1\leq j < n \leq i \leq N-1} (i+1-j)\right) \left(\prod_{n \leq j < i \leq N-1} (i-j)\right) \\
  &=& \prod_{i=1}^{n-2} i! \prod_{i=n}^{N-1} \frac{i!}{(i-n+1)!} \prod_{i=n+1}^{N-1} (i-n)! = \frac{\prod_{i=1}^{N-1} i!}{(n-1)!(N-n)!}.
\end{eqnarray}

Wrapping up the above derivations, one obtains successively
\begin{eqnarray}
\det(\mathbf{g}^{(n)}) &=&  \left( \prod_{i=1}^{N-1} i! \right)^2 (-1)^{(N-1)(N-2)/2} \frac{1}{\left[(n-1)!\right]^2 (N-n)!}, \\
\det\left[\tilde{f}_{i,n}^{(0)}(0);\ldots;\tilde{f}_{i,n}^{(N-2)}(0) \right] & = & \left( \prod_{i=1}^{N-1} i! \right)^2  \frac{(-1)^{n+1} \gamma^{n-N^2+N-1}}{\left[(n-1)!\right]^2 (N-n)!},\\
\det\left[f_i^{(0)}(0);\ldots;f_i^{(N-2)}(0); f_i(x) \right] &=& \sum_{n=1}^{N} (-1)^{1-N} f_n(x) \left( \prod_{i=1}^{N-1} i! \right)^2  \frac{\gamma^{n-N^2+N-1}}{\left[(n-1)!\right]^2 (N-n)!}.
\end{eqnarray}
Finally, we obtain
\begin{equation}
  P_{x}(x) =  -\sum_{n=1}^{N}  f_n(x) \frac{\gamma^{N+n-1}}{\left[(n-1)!\right]^2 (N-n)!},
\end{equation}
where $\gamma = -\frac{N}{E_0}$.\\

\begin{figure}[htb]
\centering
\subfigure[PDF of $x=||\mathbf{H}||^2_F$ ]{\epsfig{figure=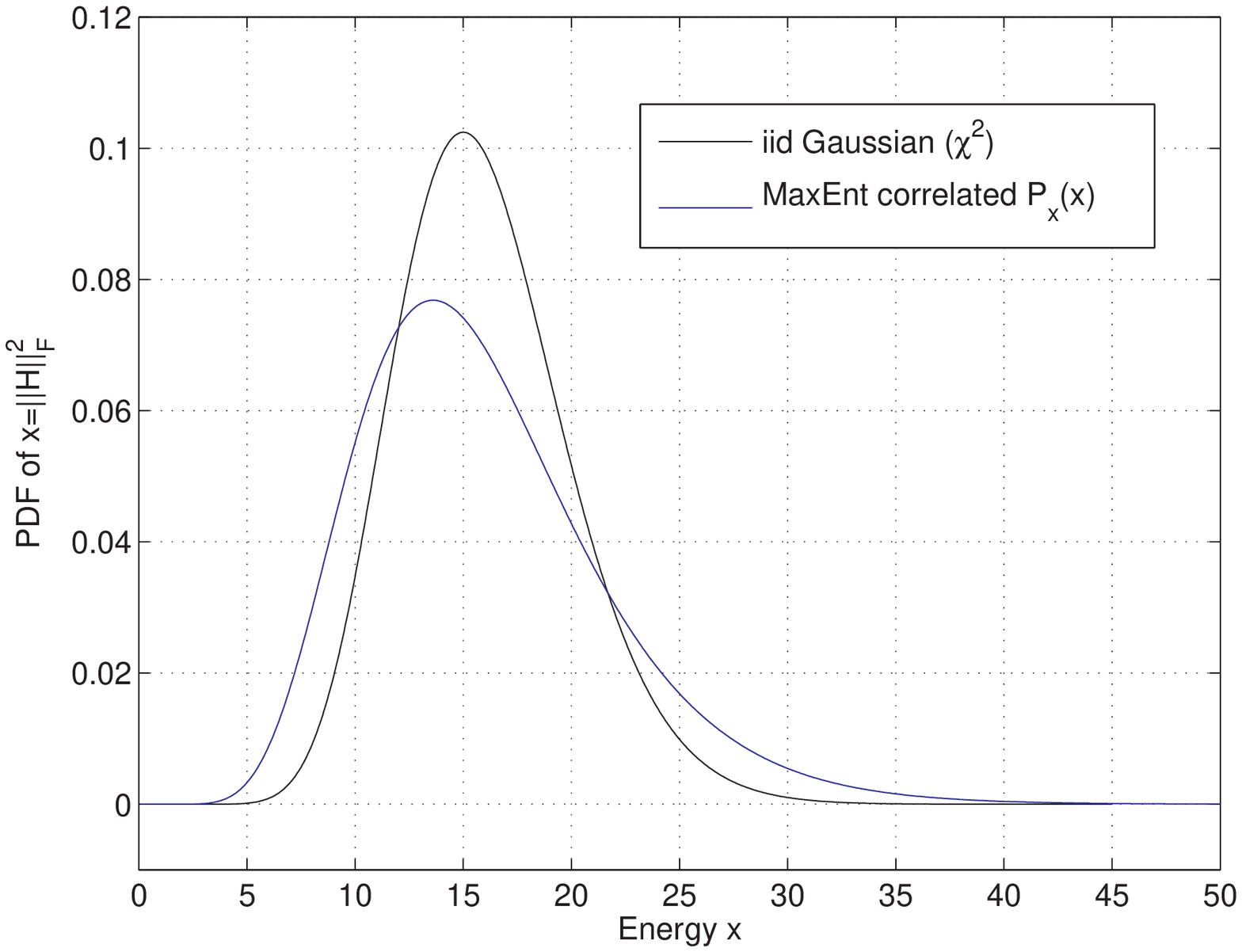,height=6.2cm}\label{pdfs_4x4}}
\subfigure[CDF of instantaneous mutual information $I$]{\epsfig{figure=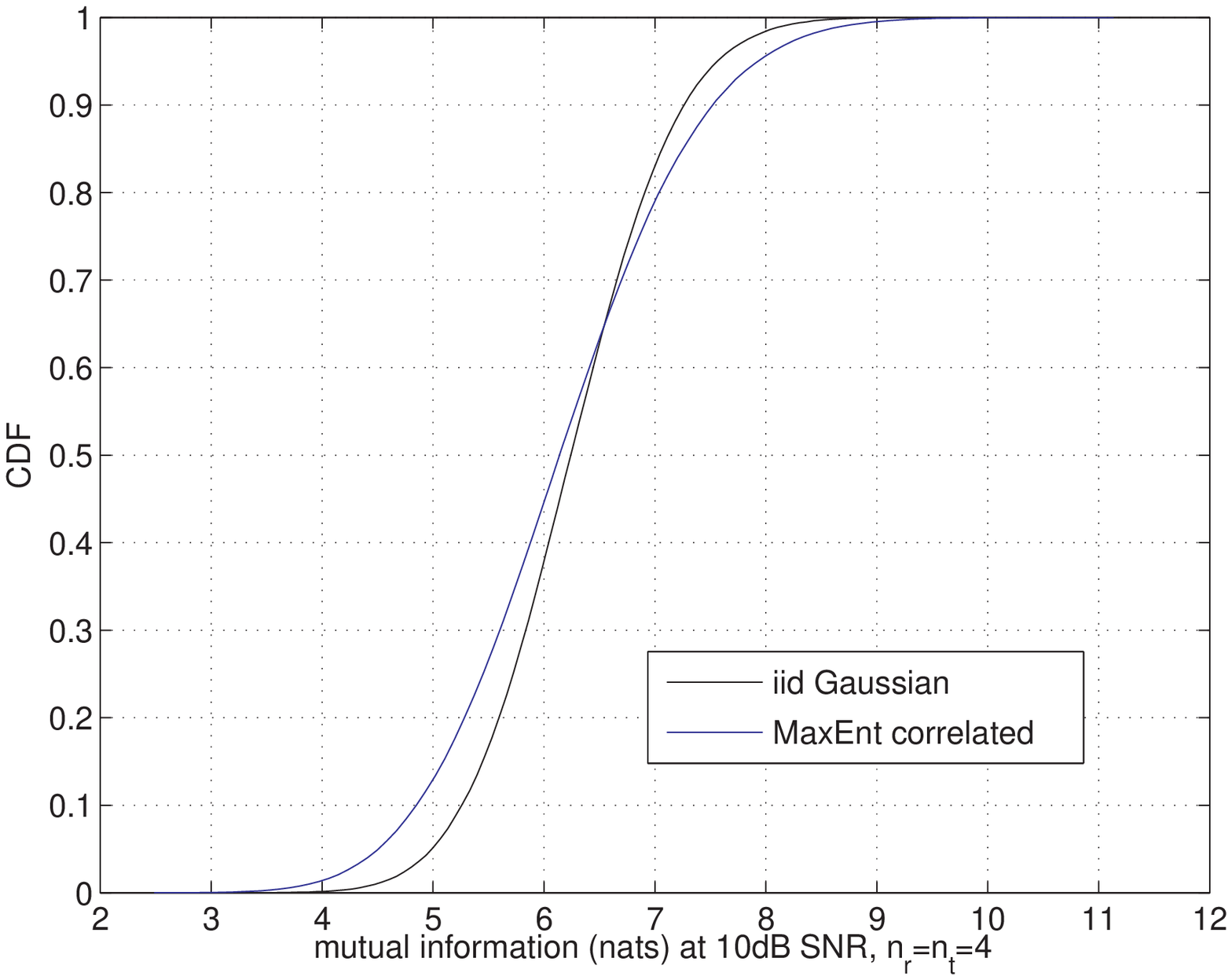,height=6.2cm}\label{Icdf_4x4}}
\caption{Amplitude and mutual information distributions of the proposed channel models for a $4\times 4$ antennas setting.}
\end{figure}

The corresponding PDF is shown in Figure~\ref{pdfs_4x4}, as well as the PDF of the instantaneous power of a Gaussian i.i.d. channel of the same size and mean power. As expected, the energy distribution of the proposed model is more spread out than the energy of a Gaussian i.i.d. channel. 

Figure~\ref{Icdf_4x4} shows the CDF curves of the instantaneous mutual information achieved over the channel described in eq.~(\ref{awgn_channelmodel}) by these two channel models. The proposed model differs in particular in the tails of the distribution: for instance, the 1\% outage capacity is reduced from 4.5 to 3.9 nats w.r.t. the Gaussian i.i.d. model.

\subsection{Limited-rank covariance matrix}

In this section, we address the situation where the modeler takes into account the existence of a covariance matrix of rank $L < N$ (we assume that L is known). 
As in the full-rank case, we will use the eigendecomposition $\mathbf{Q} = \mathbf{U}\Lambda\mathbf{U}^H$ of the covariance matrix, with $\Lambda = \diag(\lambda_1, \ldots, \lambda_L, 0, \ldots, 0)$. Let us denote $\Lambda_L = \diag(\lambda_1, \ldots, \lambda_L)$.
The maximum entropy probability density of $\mathbf{Q}$ with the extra rank constraint is unsurprisingly similar to the one derived in Section~\ref{sectionmaxentQ}, with the difference that all the energy is carried by the first $L$ eigenvalues, \emph{i.e.} $\mathbf{U}$ is uniformly distributed over $\mathcal{U}(N)$, while
\begin{equation}
  P_{\Lambda_L}(\Lambda_L) = \left(\frac{L^2}{N E_0} \right)^{L^2} \prod_{n=1}^L \frac{1}{n! (n-1)!} \exp\left(-\frac{L^2}{N E_0} \sum_{i=1\ldots L} \lambda_i\right) \prod_{i<j\leq L} (\lambda_i-\lambda_j)^2 \label{finalpdflambdaL}.\\
\end{equation}

However, the definition of the conditional probability density $P_{\mathbf{H}|\mathbf{Q}}(\v{h},\mathbf{U},\Lambda)$  in eq.~(\ref{maxent_G_corr}) does not hold when $\mathbf{Q}$ is not full rank: $\mathbf{h}$ becomes a degenerate Gaussian random variable. Its projection 
in the L-dimensional subspace associated to the non-zero eigenvalues of $\mathbf{Q}$ follows a Gaussian law, whereas the probability of $\v{h}$ being outside of this subspace is zero. 
The conditional probability in eq.~(\ref{conditional_H_Q}) must therefore be rewritten as
\begin{equation}
  P_{\mathbf{H}|\mathbf{Q}}(\v{h},\mathbf{U}, \Lambda_L) = \mathbbm{1}_{\{ \v{h} \in \mathrm{Span}(\mathbf{U}_{[L]}) \} } \frac{1}{\pi^L \prod_{i=1}^L \lambda_i} \e^{-\v{h}^H \mathbf{U}_{[L]}\Lambda_L^{-1}{\mathbf{U}_{[L]}}^H \v{h} }, \label{conditional_H_Q_limitedrank}
\end{equation}
where $\mathbf{U}_{[L]}$ denotes the $N \times L$ matrix obtained by truncating the last $N-L$ columns of $\mathbf{U}$. The indicator function ($\mathbbm{1}_{\mathcal{A}}=1$ if statement $\mathcal{A}$ is true, 0 else) ensures that $P_{\mathbf{H}|\mathbf{Q}}(\v{h},\mathbf{U}, \Lambda)$ is zero for $\v{h}$ outside of the column span of $\mathbf{U}_{[L]}$.\\

We need now to marginalize $\mathbf{U}$ and $\Lambda$ in order to obtain the PDF of $\v{h}$:
\begin{equation}
  P_{\mathbf{H}}(\v{h}) = \int_{\mathcal{U}(N) \times{\mathbb{R}^+}^L}  P_{\mathbf{H}|\mathbf{Q}}(\v{h},\mathbf{U},\Lambda_L)  P_{\Lambda_L}(\Lambda_L) \d\mathbf{U}\d\Lambda_L. 
\end{equation}
However, the expression of $P_{\mathbf{H}|\mathbf{Q}}(\v{h},\mathbf{U}, \Lambda_L)$ does not lend itself directly to the marginalization described in Section~\ref{section_marginalizeQ}, since the zero eigenvalues of $\mathbf{Q}$ complicate the analysis. However, this can be avoided by performing the marginalization of the covariance in an L-dimensional subspace. In order to see this, consider an $L \times L$ unitary matrix  $\mathbf{B}_L$, and note that the $N \times N$ block-matrix $\mathbf{B} = \left(\begin{array}{cc} \mathbf{B}_L & 0 \\ 0 & \Id_{N-L} \end{array} \right)$ is unitary as well.
Since the uniform distribution over $\mathcal{U}(N)$ is unitarily invariant, $\mathbf{U} \mathbf{B}$ is uniformly distributed over $\mathcal{U}(N)$, and for any $\mathbf{B}_L \in \mathcal{U}(L)$ we have
\begin{equation}
  P_{\mathbf{H}}(\v{h}) = \int_{\mathcal{U}(N) \times{\mathbb{R}^+}^L}  P_{\mathbf{H}|\mathbf{Q}}(\v{h},\mathbf{U} \mathbf{B},\Lambda_L)  P_{\Lambda_L}(\Lambda_L) \d\mathbf{U}\d\Lambda_L. 
\end{equation}
Furthermore, since $\int_{\mathcal{U}(L)} \d\mathbf{B}_L = 1$,
\begin{eqnarray}
  P_{\mathbf{H}}(\v{h}) &=& \int_{\mathcal{U}(L)} \int_{\mathcal{U}(N) \times{\mathbb{R}^+}^L}  P_{\mathbf{H}|\mathbf{Q}}(\v{h},\mathbf{U} \mathbf{B},\Lambda_L)  P_{\Lambda_L}(\Lambda_L) \d\mathbf{U}\d\Lambda_L \d\mathbf{B}_L \\
  &=&  \int_{\mathbf{U} \in \mathcal{U}(N)}  \mathbbm{1}_{\{ \v{h} \in \mathrm{Span}(\mathbf{U}_{[L]}) \} }  \int_{\mathcal{U}(L)\times{\mathbb{R}^+}^L} \frac{1}{\pi^L \prod_{i=1}^L \lambda_i} \e^{-\v{h}^H \mathbf{U}_{[L]}\mathbf{B}_L\Lambda_L^{-1}{\mathbf{B}_L^H \mathbf{U}_{[L]}}^H \v{h} }  P_{\Lambda_L}(\Lambda_L) \d\mathbf{B}_L \d\Lambda_L \d\mathbf{U} \\
  & = &  \int_{\mathbf{U} \in \mathcal{U}(N)}  \mathbbm{1}_{\{ \v{h} \in \mathrm{Span}(\mathbf{U}_{[L]}) \} } P_{\v{k}}({\mathbf{U}_{[L]}}^H \v{h}) \d\mathbf{U}, \label{integralk_intermed0}
\end{eqnarray}
where (\ref{integralk_intermed0}) is obtained by letting $\v{k} = {\mathbf{U}_{[L]}}^H \v{h}$ and 
\begin{equation}
P_{\v{k}}(\v{k}) = \int_{\mathcal{U}(L)\times{\mathbb{R}^+}^L} \frac{1}{\pi^L \prod_{i=1}^L \lambda_i} \e^{-\v{k}^H \mathbf{B}_L\Lambda_L^{-1}\mathbf{B}_L^H \v{k} }  P_{\Lambda_L}(\Lambda_L) \d\mathbf{B}_L \d\Lambda_L. \label{integralk_intermed}
\end{equation}
We can then exploit the similarity of eqs.~(\ref{integralk_intermed}) and (\ref{integralH_intermed1}), and, by the same reasoning as in Section~\ref{section_marginalizeQ}, conclude directly that $\v{k}$ is isotropically distributed in $\mathcal{U}(L)$, and that its PDF depends only on its Frobenius norm, following
\begin{equation}
  P_{\v{k}}(\v{k})  = \frac{1}{S_L(\v{k}^H \v{k})} P^{(L)}_{x}(\v{k}^H \v{k}), 
\end{equation}
where
\begin{equation}
  P^{(L)}_{x}(x) =  \frac{2}{x}\sum_{i=1}^{L}  \left(-L \sqrt{\frac{x}{N E_0}}\right)^{L+i} K_{i+L-2}\left(2 L\sqrt{\frac{x}{N E_0}}\right) \frac{1}{\left[(i-1)!\right]^2 (L-i)!}.
\end{equation}
Finally, note that $\v{h}^H\v{h}=\v{k}^H\v{k}$, and that the marginalization over the random rotation that transforms $\v{k}$ into $\v{h}$ in eq.~(\ref{integralk_intermed0}) preserves the isotropic property of the distribution. Therefore, 
\begin{equation}
  P_{\v{h}}(\v{h})  = \frac{1}{S_N(\v{h}^H \v{h})} P^{(L)}_{x}(\v{h}^H \v{h}).
\end{equation} \\

\begin{figure}[htb]
\centering
\epsfig{figure=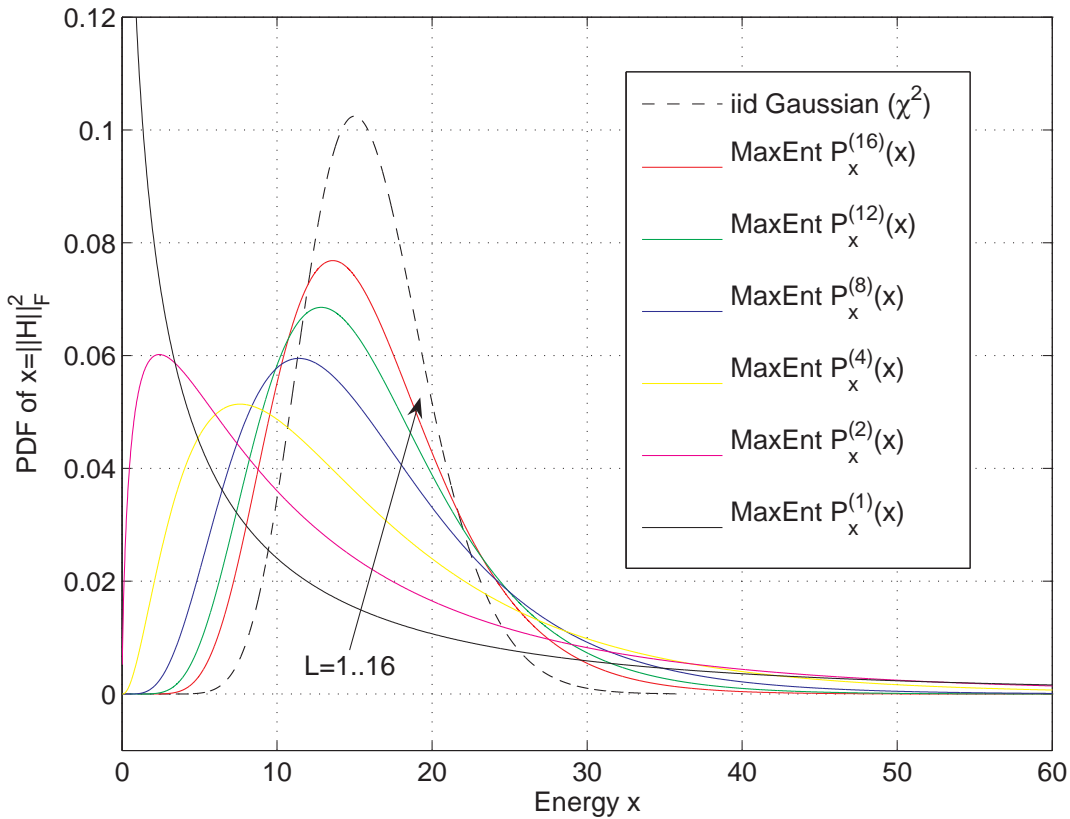,height=11cm}
\caption{Examples of the limited-rank covariance distribution $P^{(L)}_{x_N}(x)$ for $L=1, 2, 4, 8, 12$ and 16, and $\chi^2$ with 16 degrees of freedom, for $N E_0 = 16$.}
\label{rank_L_PDF_fig}
\end{figure}

Examples of the corresponding PDFs for $L=1, 2, 4, 8, 12$ and 16 are represented on Fig.~\ref{rank_L_PDF_fig} for a $4 \times 4$ channel ($N=16$), together with the PDF of the instantaneous power of a Gaussian i.i.d. channel of the same size and mean power. 
As expected, the energy distribution of the proposed MaxEnt model is more spread out than the energy of a Gaussian i.i.d. channel. \\ 

\begin{figure}[htb]
\centering
\epsfig{figure=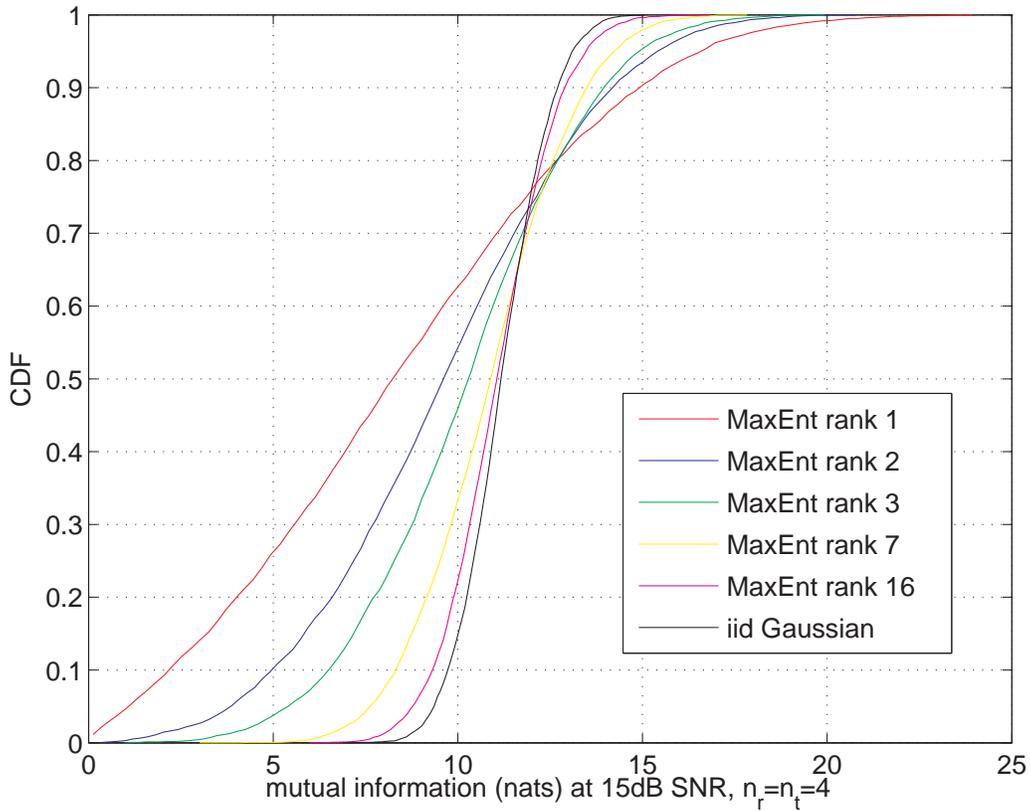,height=11cm}
\caption{CDF of the instantaneous mutual information of a $4\times 4$ flat-fading channel for the MaxEnt model with various covariance ranks, at 15dB SNR.}
\label{MI_rankL_CDF_fig}
\end{figure}

The CDF of the mutual information achieved over the limited-rank ($L<16$) and full rank ($L=16$) covariance MaxEnt channel at a SNR of 15 dB is pictured on Figure~\ref{MI_rankL_CDF_fig} for various ranks $L$, together with the Gaussian i.i.d. channel.
The proposed model differs in particular in the tails of the distribution. In particular, the outage capacity for low outage probability is greatly reduced w.r.t. the Gaussian i.i.d. channel model.

\section{Conclusion}
\label{section_conclusion}

In this paper, the maximum entropy principle is used to derive several models of wireless flat-fading channels for various cases of a priori knowledge about the channel properties. First, the cases of average channel energy and known upper-bound on the channel energy were studied. 
Subsequently, the issue of taking into account an unknown amount of spatial correlation in MIMO channel models was addressed. The entropy maximizing distribution of the covariance matrix under a average trace constraint was shown to be a Wishart distribution, and the corresponding probability density function of the channel matrix was shown to be described analytically by a function of the channel Frobenius norm. This model was generalized to the case where the covariance matrix is rank-deficient with known rank.
The proposed channel models were compared to the commonly used Gaussian i.i.d. models in terms of the statistics of the achieved mutual information for a given noise level. The proposed models exhibit slightly lower average mutual information, in line with the rule that channel correlation decreases its capacity, and a higher variance than the Gaussian i.i.d. model, which reflects the presence of shadow fading.

\section*{Acknowledgements}
This research was supported by France Telecom and by the NEWCOM network of excellence. The authors would like to thank an anonymous reviewer for pointing out reference \cite{Adhikari_maxent_elastodynamics2006}.

\bibliographystyle{IEEEbib}
\bibliography{biblio}

\end{document}